\documentclass[prl,twocolumn,superscriptaddress]{revtex4}
\usepackage{amssymb}

\usepackage{dcolumn}
\usepackage{epsfig}
\usepackage{amsmath}

\begin{document}

\title{Two-gap superconductivity in MgB$_{2}$: clean or dirty?}
\author{I. I. Mazin}
\affiliation{Center for Computational Materials Science, Naval Research Laboratory,
Washington, DC 20375-5000, USA}
\author{O. K. Andersen}
\author{O. Jepsen}
\author{O. V. Dolgov}
\author{J. Kortus}
\affiliation{Max-Planck-Institut f{\"u}r Festk{\"o}rperforschung, Heisenbergstr. 1,
D-70569 Stuttgart, Germany}
\author{A. A. Golubov}
\affiliation{University of Twente, Department of Applied Physics, NL-7500 AE Enschede,
The Netherlands}
\author{A. B. Kuz'menko}
\author{D. van der Marel}
\affiliation{University of Groningen, Nijenborgh 4, 9747 AG Groningen, The Netherlands}
\date{\today}

\begin{abstract}
A large number of experimental facts and theoretical arguments favor a
two-gap model for superconductivity in MgB$_{2}$. However, this model
predicts strong suppression of the critical temperature by interband
impurity scattering and, presumably, a strong correlation between the
critical temperature and the residual resistivity. No such correlation has
been observed. We argue that this fact can be understood if the band
disparity of the electronic structure is taken into account, not only in the
superconducting state, but also in normal transport.
\end{abstract}

\maketitle

\begin{figure}[tbp]
\epsfig{file=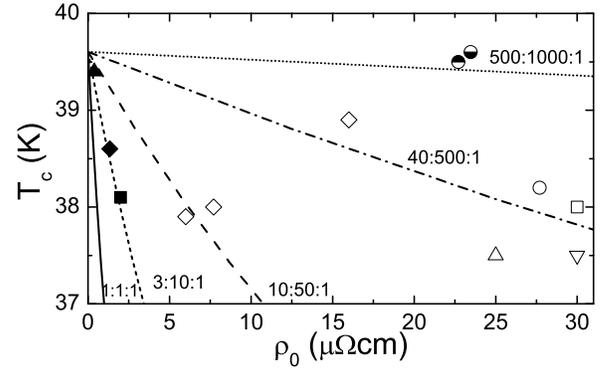,width=.9\linewidth,clip=true}
\caption{ Critical temperature for samples of varying quality as a function
of the residual resistivity. The theoretical curves are computed in the
two-band model, according to Ref.\protect\cite{golubov}, with different
ratios: $\Gamma _{\protect\sigma \protect\sigma }\left/ N_{\protect\sigma %
}\left( 0\right) \right. :\Gamma _{\protect\pi \protect\pi }\left/ N_{%
\protect\pi }\left( 0\right) \right. :\Gamma _{\protect\sigma \protect\pi %
}\left/ N_{\protect\pi }\left( 0\right) \right. .$ Filled symbols refer to
`high-quality samples': dense wires ($\blacktriangle $)\protect\cite{wires}
and single crystals ($\blacklozenge ,\blacksquare )$ \protect\cite{xtals,Xtals}.
Half-filled symbols refer to `high-$T_{c}$, high-$\protect\rho $' samples %
\protect\cite{Tu,kuzm}. Open symbols refer to samples of intermediate
quality ($\lozenge ,\circ ,\triangle ,\triangledown ,\square $) \protect\cite%
{other-rho}. }
\label{fig:rhotc}
\end{figure}

Most researchers ascribe the superconductivity in MgB$_{2}$ \cite{akimitsu}
to the electron-phonon mechanism, enhanced by interband anisotropy of the
order parameter \cite{liu,choi}. Interband anisotropy, as expressed by the
two-gap model \cite{liu,shulga}, offers a simple explanation of many
anomalous experimental findings, most importantly of tunneling and
thermodynamic measurements \cite{review}. But there is a strong argument
against it: As illustrated in Fig. \ref{fig:rhotc}, existing bulk samples of
MgB$_{2}$ have essentially the \emph{same} critical temperature although
their residual resistivities, $\rho _{0},$ vary greatly, between 0.4 and 40 $%
\mu \Omega $ cm. Such a behavior is expected for $s$-wave pairing
(Anderson's theorem), but not when \emph{two} gaps are present. In that case
one expects $T_{c}$ to fall with increasing $\rho _{0}.$ Indeed, impurity
interband scattering (magnetic and nonmagnetic) with rate $\gamma _{\mathrm{%
\ inter}}$ suppresses two-band superconductivity as: $\Delta T_{c}\propto
\gamma _{\mathrm{inter}}\left/ \left( \pi T_{c}\right) \right. $ \cite%
{golubov}, and it is tempting to assume that $\gamma _{\mathrm{intra}}\sim
\gamma _{\mathrm{inter}}\propto \rho _{0}.$ For a sample with $\rho _{0}\sim
40$ $\mu \Omega $ cm it seems unlikely that $\gamma _{\mathrm{inter}}$ can
be smaller than $\pi T_{c}$. In fact, the body of experimental evidence
(Fig. \ref{fig:rhotc}) can be reconciled with the two-gap model only if $%
\gamma _{\mathrm{inter}}\ll \gamma _{\mathrm{intra}}$. Until this paradox is
resolved, the two-gap model for superconductivity in MgB$_{2}$ cannot be
accepted, despite much compelling evidence. Two further problems are: (a)
The high-temperature slope of the resistivity is clearly correlated with the
residual resistivity (violation of Matthiessen's rule) \cite{review}, and
(b) the plasma frequency estimated from the measured infrared reflectivity
is 5 times smaller than the calculated one \cite{kaindl,Tu,kuzm}.

In this letter we shall show that the paradox can be resolved to support the
two-gap model. It turns out that due to the particular electronic structure
of MgB$_{2},$ the impurity scattering \emph{between} the $\sigma $- and $\pi 
$-bands is exceptionally small. Thus, the large variation of the residual
resistivities reflects primarily a large variation of the scattering rate 
\emph{inside} the $\sigma $- and the $\pi $-bands, while the interband $%
\sigma \pi $-scattering plays \emph{no} role in normal transport. In the
superconducting state, the two different gaps in the $\sigma $- and the $\pi 
$-bands are preserved even in dirty samples due to the extreme weakness of
the $\sigma \pi $-interband impurity scattering.

\begin{figure}[tbp]
\epsfig{file=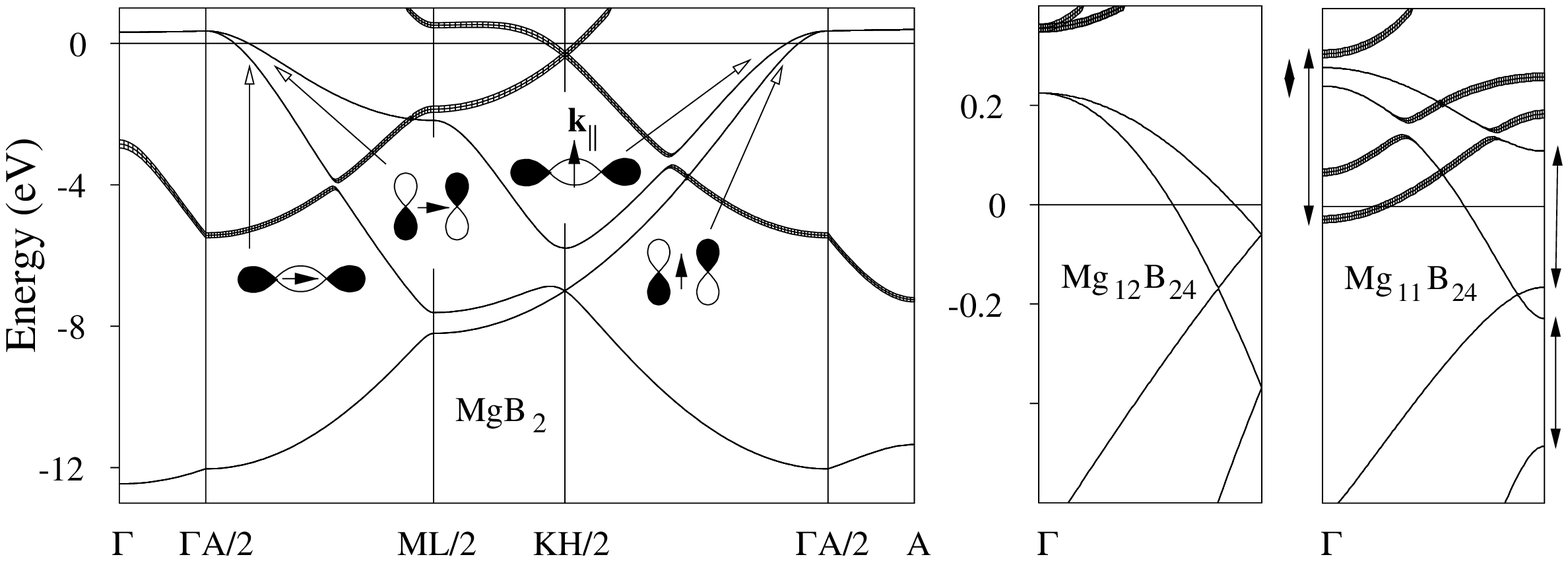,width=\linewidth}
\caption{LMTO bandstructure of MgB$_{2}$ along the $\Gamma $A-line and in
the plane $\left( k_{z}\mathrm{=}\frac{\protect\pi }{2c}\right) $ between
the $\Gamma $MK and ALH-planes, where the $\protect\sigma $ and $\protect\pi 
$ bands (fat) hybridize most. The $\Gamma $M/AL-direction is along, and the $%
\Gamma $K/AH-direction is perpendicular to a B-B bond. The orbital
characters of the heavy and light $\protect\sigma $-bands are explained in
the text. $6\times 2\times 1$ supercell bands for Mg$_{12}$B$_{24}$ and Mg$%
_{11}$B$_{24}$ are shown along the main folding-direction, $\Gamma $M. For Mg%
$_{11}$B$_{24},$ two extra electrons and protrone dded, and the nuclear charge 
of each Mg increased by 2/11,
to preserve the band filling and electroneutrality.}
\label{fig:bands}
\end{figure}

MgB$_{2}$ has two $\pi $ and three $\sigma $-bands (Fig. \ref{fig:bands})
formed by, respectively, the two B $p_{z}$ and the three bond-orbitals per
cell, or, more correctly, by the corresponding Wannier-like functions. A bond
orbital is the bonding linear combination of the two B $sp^{2}$-hybrids
which are directed along a B-B bond. The attractive potential from the Mg$%
^{2+}$ ions in the hollows between the hexagonal boron layers is felt much
stronger by a $p_{z}$-electron than by a bond-electron and, as a result, the 
$\pi $-band is pulled so far down in energy that $\sim $0.17 holes are left
at the top of the $\sigma $-band. The strong coupling of these holes to the
optical bond-stretching modes \cite{kong} is what drives the
superconductivity. Since the top of the $\sigma $-band is at $\mathbf{k}%
_{\shortparallel }\equiv \left( k_{x},k_{y}\right) \mathbf{=0}$ and is
doubly degenerate, the holes are distributed in an upper \emph{heavy} and a
lower \emph{light} band.

The basic reason why $\sigma \pi $-impurity scattering is small is that the 
$\sigma $ and $\pi $-bands are formed from different local orbitals, and
therefore are orthogonal on the \emph{atomic} scale, rather than merely on
an intermediate scale because of Bloch factors. Moreover, the layered
structure and the compactness of the B $2s$ and $2p$ orbitals makes the $%
\sigma \pi $-disparity in MgB$_{2}$ much stronger than, say, the $sd$%
-disparity in a transition metal, where the $sd$-hybridization gap is almost
as large as the $d$-bandwidth. 

Specifically, since a $p_{z}$-orbital has odd-parity, and a 
bond-orbital has even parity with
respect to the B-layer, the only route for $\sigma \pi $-hybridization is
via interlayer hopping, from a $p_{z}$-orbital in one layer
 to a bond orbital in
another layer. The corresponding hopping integral, $t_{bz},$ is, essentially,
the geometrical average of the integrals $t_{bb}^{\perp }\sim 0.1$ eV and $%
t_{zz}^{\perp }\sim 1$ eV, responsible for the $k_{z}$-dispersions of the $%
\sigma $ and $\pi $-bands \cite{kong}, and therefore small. Two further
factors limit $\sigma \pi $-coupling: The second is that, in its interaction
with the nearest bond-orbitals in the next layer, the B $p_{z}$-orbital
picks up merely the \emph{axial} projection, which is essentially the 
$s$\emph{-character}, on the boron above (or below) it. Near the top of the $%
\sigma $-band, the linear combinations of the three bond orbitals are,
however, such that the contributions from the B $s$-orbitals \emph{cancel,}
so that the top of the $\sigma $-band is purely B $p_{x},p_{y}$-like. Hence,
the only source of B $s$-character is tails of B $p$-orbitals centered at
other sites. It turns out that the wavefunctions for the heavy and light
holes $\left( \nu =h,l\right) $ are: $\left| \sigma _{\nu },\mathbf{k}%
\right\rangle \propto \sum_{\mathbf{T}}\left[ p_{\nu }\left( \mathbf{r+\tau
-T}\right) -p_{\nu }\left( \mathbf{r-\tau -T}\right) \right] e^{i\mathbf{%
k\cdot T}},$ where $\mathbf{T}$ are the lattice translations, $\pm \mathbf{%
\tau }$ are the positions of the two borons in the cell (i.e., in a bond),
and $p_{h/l}\left( \mathbf{r}\right) $ is a B $p$-orbital directed
transverse/longitudinal to the $\mathbf{k}_{\shortparallel }$-vector. From
this representation, illustrated in Fig. \ref{fig:bands}, it may be realized
that the B $s$-character often vanishes completely, and that it generally
vanishes proportional to $k_{\shortparallel }^{2}$ for the heavy-holes, and
proportional to $k_{\shortparallel }$ for the light holes. The third
limiting factor is the matching of the phase, $\varphi ,$ between the two $p_{z}
$-orbitals in a bond,  $\left| \pi _{\mp },\mathbf{k}\right\rangle \propto
\sum_{\mathbf{T}}\left[ p_{z}\left( \mathbf{r+\tau -T}\right) e^{i\varphi
\left( \mathbf{k}_{\shortparallel }\right) }\mp p_{z}\left( \mathbf{r-\tau -T%
}\right) \right] e^{i\mathbf{k\cdot T}},$ and the phase between the
corresponding B $s$-characters arising from the antibonding combination, $%
p_{n}\left( \mathbf{r+\tau }\right) -p_{n}\left( \mathbf{r-\tau }\right) .$
In the nearest-neighbor orthogonal tight-binding model for the $\pi $-bands, 
$\varphi \left( \mathbf{k}_{\shortparallel }\right) =\arg \left\{ 1+e^{i%
\mathbf{k\cdot a}}+e^{i\mathbf{k\cdot }\left( \mathbf{a-b}\right) }\right\} ,
$ where $\mathbf{a}$ and $\mathbf{b}$ are the primitive translations of the
layer.

Due to their even/odd parity, the $\sigma $ and $\pi $-bands can only
hybridize when $k_{z}\neq \frac{\pi }{c}\times $integer$.$ Even then, as
seen in Fig. \ref{fig:bands}, the $\pi _{+}$-band neither hybridizes with
the heavy $\sigma $-band when $\mathbf{k}_{\shortparallel }$ is along a
bond, nor with the light $\sigma $-band when $\mathbf{k}_{\shortparallel }$
is perpendicular to a bond. As may be realized from the pictures of the $%
\sigma $-orbitals (Fig. \ref{fig:bands}),
 the crossing with the heavy band occurs because the B $s$%
-character of that band vanishes exactly along this $\mathbf{k}$-line, and
the crossing with the light band occurs because, along that $\mathbf{k}$%
-line, the B $s$-character is purely antibonding between two borons, whereas
the $\pi _{+}$-band is purely bonding $\left( \varphi =0\right) .$ The two $%
\sigma \pi $-gaps seen in the figure are 0.2--0.3 eV, \textit{i.e., }the
hybridization matrix elements, $|\left\langle \sigma \mathbf{k}|H|\pi 
\mathbf{k}\right\rangle |,$ are merely a per cent of the $\sigma $ and $\pi $
bandwidths!

We now discuss impurity scattering and use \cite{allen}: 
\begin{equation}
\Gamma _{nn^{\prime }}=\frac{2}{\hbar N_{n}\left( 0\right) }\sum_{\mathbf{kk}%
^{\prime }}\delta (\varepsilon _{n\mathbf{k}})|\left\langle n\mathbf{k}%
|V|n^{\prime }\mathbf{k}^{\prime }\right\rangle |^{2}\delta (\varepsilon
_{n^{\prime }\mathbf{k}^{\prime }}),  \label{e1}
\end{equation}%
for the rate of scattering to band $n^{\prime }$ of an electron in band $n,$
by a {weak} localized impurity potential, $V\left( \mathbf{r}\right) .$
Here, $\sum_{\mathbf{k}}$ denotes the average over the Brillouin zone, $%
\varepsilon _{n\mathbf{k}}$ is the band energy with respect to the Fermi
level, and $N(0)=\sum_{n}N_{n}\left( 0\right) =\sum_{n\mathbf{k}}\delta
(\varepsilon _{n\mathbf{k}})$ is the density of states per spin and cell.
Typical defects for MgB$_{2}$ are Mg-vacancies and Mg-substitutional
impurities, which form easily, and B-site substitutions like N and C, which
have a higher energy cost. The potential $V\left( \mathbf{r}\right) $ for a
localized Mg-defect has the full point-symmetry of the site and, like the Mg$%
^{2+}$potential in the crystal, is felt more by a $p_{z}$-orbital than by a
bond orbital. Hence, the largest matrix elements are those involving $p_{z}$%
-orbitals near the impurity, i.e., the largest perturbation is of the
energies of the $p_{z}$-orbitals on the B hexagons immediately above and
below the impurity, and of the corresponding $t_{zz}^{\perp }.$ This means
that $\Gamma _{\pi \pi }$ should be large. Screening perturbs the energies
of the bond orbitals surrounding the impurity, and also perturbs $%
t_{bb}^{\perp },$ but to a lesser extent. Hence, we expect that $\Gamma
_{\pi \pi }>\Gamma _{\sigma \sigma }$ for Mg-defects, albeit not for B-site
substitutions. What contributes to $\Gamma _{\sigma \pi },$ are matrix
elements involving a $p_{z}$ and a bond-orbital, and most importantly, those
on either side of a Mg-defect. Since this matrix element is the perturbation
of $t_{zb},$ it is expected to be intermediate between those of $%
t_{zz}^{\perp }$ and $t_{bb}^{\perp },$ like for the $\sigma \pi $%
-hybridization. Moreover, since the impurity potential is fairly constant
around a neighboring boron, a $p_{z}$-orbital still picks up merely the B $s$%
-character which vanishes as $k_{\shortparallel }^{2}$ for the heavy and as $%
k_{\shortparallel }$ for the light holes. This makes $|\left\langle \sigma 
\mathbf{k}|V|\pi \mathbf{k}^{\prime }\right\rangle |$ minute because $k_{Fh}$
and $k_{Fl}$ are very small. Also the mismatch of phases between the $\sigma 
$ and $\pi $-functions will tend to reduce $|\left\langle \sigma \mathbf{k}%
|V|\pi \mathbf{k}^{\prime }\right\rangle |.$ Finally, squaring this small
matrix element and inserting it in (\ref{e1}), leads to an exceedingly small 
$\Gamma _{\sigma \pi }.$

To gain quantitative understanding of the disparity between the scattering
rates we have performed LMTO supercell calculations for various impurities.
Since the induced $\sigma \pi $-gaps, $2|\left\langle \sigma \mathbf{k}%
|V|\pi \mathbf{k}^{\prime }\right\rangle |,$ are sensitive to their position
within the $\sigma $-band (the B $s$-factor), we must choose a supercell
which provides band-foldings near $\varepsilon _{F}.$ The results shown in
Fig. \ref{fig:bands} were obtained with a $6\times 2\times 1$ supercell. The
bands labeled Mg$_{12}$B$_{24}$ are the same as those in the left panel, but
folded into the smaller zone. The heavy $\sigma $-band now crosses itself
closely below $\varepsilon _{F},$ while the heavy-light and light-light
crossings are a bit further down. The $\pi _{-}$-band (fat) slightly above
the top of the $\sigma $-band was originally at ML/2 and has been folded 3
times into $\Gamma .$ The Mg$_{11}$B$_{24}$ bands illustrate the effects of
a Mg-vacancy: While the three $\pi $-bands get split by 0.35 eV, and the
heavy and light $\sigma $-bands by 0.27 eV (but by 0.04 eV at $\Gamma )$,
the $\sigma \pi $-splitting of the heavy band is merely 0.015 eV and that of
the light band is merely 0.030 eV! The squares of these splittings give
estimates for the corresponding $\Gamma $'s. For Mg-vacancies therefore, 
\begin{equation}
\Gamma _{\pi \pi }>\Gamma _{\sigma \sigma }\gg \Gamma _{\sigma \pi }.
\label{e5}
\end{equation}%
We found very similar results for systems in which the Mg-vacancy was
compensated by substitution of B by two C or one N: For Mg$_{15}$B$_{31}$N,
the $\pi \pi $-splitting was 0.4 eV, the $\sigma \sigma $-splitting 0.3 eV,
and the $\sigma \pi $-splittings less than 0.03 eV.

Let us now investigate how the relation (\ref{e5}) infuences the transport properties.
These depend both on the impurity scattering and on the electron-phonon
interaction (EPI). The interband anisotropy should be taken into account
both in the impurity scattering (as outlined above), and in the EPI. The
latter can be characterized by two sets of four spectral functions each: the
standard Eliashberg functions $\alpha ^{2}F_{nn^{\prime }}(\omega )$, which
define the superconducting properties and thermodynamical properties like
the electronic specific heat and the de Haas-van Alphen mass
renormalizations, and the transport Eliashberg functions $\alpha _{\mathrm{tr%
}}^{2}F_{nn^{\prime }}(\omega )$. Of the calculated $\alpha
^{2}F_{nn^{\prime }}\left( \omega \right) $ functions \cite{Gol-heat} (the
details of the calculations are as in Ref. \cite{kong}), $\alpha
^{2}F_{\sigma \sigma }(\omega )$ exhibits a large peak at $\omega \approx 70$
meV. Defining $\lambda _{nn^{\prime }}=2\int \omega ^{-1}\alpha
^{2}F_{nn^{\prime }}(\omega )d\omega $, we obtain the partial EPI constants,
shown in Table \ref{tab:lambda}, which are similar to those obtained in \cite%
{liu}. In the following we assume that $\Gamma _{\mathrm{inter}}=0$, so the
clean limit is appropriate ($T_{c}$ is independent of the intraband $\Gamma $%
's). The superconducting properties in the clean two-band model have been
investigated in detail \cite{choi,Gol-heat}. Therefore we shall now
concentrate on the normal transport.

The explicit expression for the conductivity in the two-band model is \cite%
{allenrho} {(omitting Cartesian indices)} 
\begin{eqnarray}
1/\rho _{\mathrm{DC}}(T) &=&\frac{1}{4\pi }\sum_{n=\sigma ,\pi }\omega _{%
\mathrm{pl\,}\,n}^{2}\left/ W_{n}(0,T)\right. ,  \label{rho} \\
W_{\sigma }(0,T) &=&\gamma _{\sigma }+\frac{\pi }{T}\int_{0}^{\infty
}d\omega \frac{\omega }{\sinh ^{2}(\omega /2T)}  \notag \\
&\times &\left[ \alpha _{\mathrm{tr}}^{2}(\omega )F_{\sigma \sigma }(\omega
)+\alpha _{\mathrm{tr}}^{2}(\omega )F_{\sigma \pi }(\omega )\right] ,  \notag
\end{eqnarray}%
where $\gamma _{\sigma }=\gamma _{\sigma \sigma }+\gamma _{\sigma \pi }$, $%
\gamma _{\pi }=\gamma _{\pi \pi }+\gamma _{\pi \sigma }$, and $\gamma
_{nn^{\prime }}\simeq 2\Gamma _{nn^{\prime }}.$ Eq.~(\ref{rho}) is
essentially the standard parallel-conductor formula. Our assumptions are
that $\gamma _{\pi \sigma }\approx 0$ and that $\gamma _{\sigma }$ differs
much less than $\gamma _{\pi }$ between `good' (e.g., Ref.\cite{wires}) and
`bad' samples (e.g. \cite{Tu}).

The role of interband anisotropy, clearly visible in Table \ref{tab:lambda},
is different in superconductivity and electric transport. For instance, the
critical temperature is given by the maximum eigenvalue of the $\lambda $%
-matrix \cite{allen,allenrho} (\textit{i.e.,} mostly by its maximum
element), while the conductivity (\ref{rho}) is the sum of the partial
conductivities. At high temperature, therefore, the slope $d\rho /dT$ is
determined by $\sum_{n}\omega _{\mathrm{pl}\text{\thinspace }n}^{2}/\lambda
_{tr\,n},$ \textit{i.e.}, by the smallest $\lambda _{\mathrm{tr}\,n}\equiv
\sum_{n^{\prime }}\lambda _{\mathrm{tr}\,nn^{\prime }}.$

\begin{figure}[tbp]
\epsfig{file=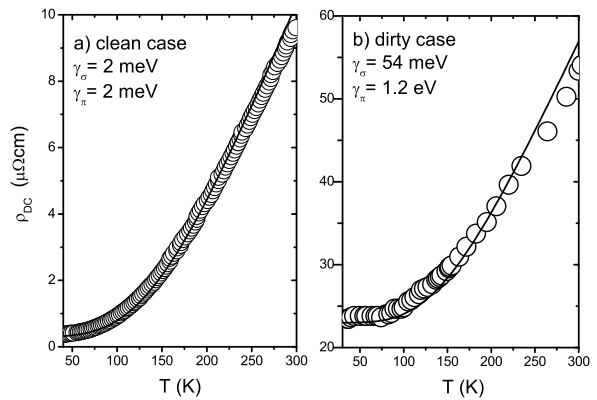,width=.9\linewidth,clip=true}
\caption{The DC resistivity in the clean (a) and dirty (b) case compared to
experimental data for dense wires \protect\cite{wires} and for c-oriented
films \protect\cite{Tu}, respectively. The lines are calculated in the
effective two-band model with the indicated scattering and \textit{ab initio}
plasma frequencies $\protect\omega _{\protect\sigma }^{ab}$= 4.14 eV, $%
\protect\omega _{\protect\pi }^{ab}$= 5.89 eV, $\protect\omega _{\protect%
\sigma }^{c}$= 0.68 eV, and $\protect\omega _{\protect\pi }^{c}$= 6.85 eV. }
\label{fig:rho}
\end{figure}
In Fig.\ref{fig:rho} we show the temperature dependence of the DC
resistivity for (a) a clean case with $\gamma _{\sigma }=\gamma _{\pi }=2$
meV, and (b) a dirty case with $\gamma _{\sigma }=54$ meV and $\gamma _{\pi
}=1.2$ eV. In the two cases, all plasma frequencies are the same. The model
is seen to describe both cases well. Note that $\gamma _{\sigma }$ and $%
\gamma _{\pi }$ determine not only the residual resistivity, but also the
temperature dependence of the resistivity. In a one-band model, it would be
impossible to reconcile the data of Refs. \cite{Tu,kuzm} with those of Ref. %
\cite{wires} if they differ only by impurity concentrations, and the
corresponding violation of Matthiessen's rule would be totally inexplicable.

Why is the temperature dependence of the resistivity so different in these
two cases? Let us compare the clean limit, $\gamma _{\sigma }=\gamma _{\pi
}=0,$ with the dirty-Mg-layer's limit, $\gamma _{\pi }=\infty $, $\gamma
_{\sigma }=0$. Of the two parallel conducting channels, in the former case
the $\pi $-bands are responsible for conductivity at high temperatures, as
was mentioned above, and even at $T\approx 0$ the conductivity is mostly due
to the $\pi $-bands, their plasma frequency being higher than that one of
the $\sigma $-bands. Since the EPI constant for the $\pi $-bands is small,
the temperature dependence of the resistivity is weak. On the contrary, in
the dirty case, the $\pi $-bands do not conduct, due to an overwhelming
impurity scattering, and the electric current is carried only by the $\sigma 
$-bands. It is the strong EPI for this band which causes the temperature
dependence of the resistivity in dirty samples..

To conclude, we suggest a new model for electric transport in MgB$_{2}$. The
main ingredients of the model are (i) \emph{interband }impurity scattering
in MgB$_{2}$ is small, even in low-quality samples; (ii) \emph{intraband}
impurity scattering in the $\sigma $-band is small relative to the \emph{%
intraband} $\pi $-band scattering; (iii) high-resistivity samples differ
from good samples mostly by the \emph{intraband} $\pi $-band scattering
rate. Of course, (iv) the phonon scattering is stronger
 in the $\sigma $-band. This
model explains well such seemingly inexplicable experimental facts as (1)
absence of direct correlation between the residual resistivity and the
critical temperature, expected in the two-gap model and (2) a strong
correlation between the residual resistivity and the slope $d\rho /dT$ in the
normal state.
 Finally, we would like to point out that the existence
of two qualitatively different scattering rates in the two bands should
manifest itself in other experiments, such as optical and microwave
spectroscopy, or Hall effect. In particular, seemingly mysterious observations
of the anomalously small Drude weight in infrared absorption\cite%
{kaindl,Tu,kuzm} are probably due to overdamping of the Drude contribution
from the $\pi $-bands, so that the observed Drude peak comes essentially
from the $\sigma $-bands. The latter have small plasma frequencies and are
additionally renormalized by the electron-phonon coupling. At the same time,
the overdamped Drude peak from the $\pi $-electrons manifest itself as a
broad background extending to high frequencies.

\acknowledgments JK would like to thank the Schloe{\ss }mann Foundation for
financial support. IIM acknowledge partial support from the Office of Naval
Research.

\begin{table}[tbp]
\begin{ruledtabular}
\begin{tabular}{ccccc}  
 & $\lambda_{\sigma\sigma}$ &  $ \lambda_{\pi\pi}$ &
             $\lambda_{\sigma\pi}$ & $
\lambda_{\pi\sigma}$
   \\  \hline
transport       & 0.80  &
0.41   & 0.30  & 0.15
   
\\
superconducting & 1.02  &  0.45   & 0.21  & 0.16
\\
\end{tabular}
\caption{Superconducting and
transport coupling constants
$\protect\lambda$
for the effective two-band
model. The partial densities
of states at the Fermi level for the
two bands
have values of
$N_{\sigma}(0)$=0.15 and
$N_{\pi}(0)$=0.21 (states/cell$\cdot$
spin$\cdot$eV).
\label{tab:lambda}
}
\end{ruledtabular}
\end{table}

\end{document}